\documentclass[11pt,a4paper]{article}
\pdfoutput=1
\usepackage{jcappub}
\usepackage{subfig}
\usepackage{caption}
\usepackage{hyperref}
\usepackage{amsmath}
\usepackage{mathtools}
\usepackage{tablefootnote}
\usepackage{soul}
\usepackage{ulem}
\usepackage{color}
\usepackage[usenames,dvipsnames,svgnames,table]{xcolor}
\usepackage{bm}
\usepackage{tikz-feynman}

\title{Measuring the Energy Scale of Inflation with Large Scale Structures}

\author[a,b]{Nicola Bellomo,}
\emailAdd{nicola.bellomo@icc.ub.edu}

\author[c,d,e]{Nicola Bartolo,}
\emailAdd{nicola.bartolo@pd.infn.it}

\author[a,f]{Raul Jimenez,}
\emailAdd{raul.jimenez@icc.ub.edu}

\author[c,d,e,g]{Sabino Matarrese,}
\emailAdd{sabino.matarrese@pd.infn.it}

\author[a,f]{Licia Verde}
\emailAdd{liciaverde@icc.ub.edu}
  
\affiliation[a]{ICC, University of Barcelona, IEEC-UB, Mart\'i  i Franqu\`es, 1, E-08028 Barcelona, Spain}
\affiliation[b]{Dept. de  F\'isica Qu\`antica i Astrof\'isica, Universitat de Barcelona, Mart\'i  i Franqu\`es 1, E-08028 Barcelona, Spain}  
\affiliation[c]{Dipartimento di Fisica e Astronomia G. Galilei, Universit\`a degli Studi di Padova, via F. Marzolo 8, I-35131, Padova, Italy.}
\affiliation[d]{INFN, Sezione di Padova, via F. Marzolo 8, I-35131, I-35131 Padova, Italy.}
\affiliation[e]{INAF - Osservatorio Astronomico di Padova, vicolo dell'Osservatorio 5, I-35122 Padova, Italy.}
\affiliation[f]{ICREA, Pg. Lluis Companys 23, Barcelona, E-08010, Spain.} 
\affiliation[g]{Gran Sasso Science Institute, viale F. Crispi 7, I-67100 L'Aquila, Italy.}

\dedicated{We dedicate this paper to the memory of our friend and colleague Bepi Tormen who did \\ pioneering work in the understanding of the abundance and clustering of dark matter halos.}

\abstract{
The determination of the inflationary energy scale represents one of the first step towards the understanding of the early Universe physics. The (very mild) non-Gaussian signals that arise from any inflation model carry information about the energy scale of inflation and may leave an imprint in some cosmological observables, for instance on the clustering of high-redshift, rare and massive collapsed structures. In particular, the graviton exchange contribution due to interactions between scalar and tensor fluctuations leaves a specific signature in the four-point function of curvature perturbations, thus on clustering properties of collapsed structures. We compute the contribution of graviton exchange on two- and three-point function of halos, showing that at large scales $k\sim 10^{-3}\ \mathrm{Mpc}^{-1}$ its magnitude is comparable or larger to that of other primordial non-Gaussian signals discussed in the literature. This provides a potential route to probe the existence of tensor fluctuations which is alternative and highly complementary to B-mode polarisation measurements of the cosmic microwave background radiation.
}

\begin{document}
\maketitle

\section{Introduction}
The inflationary paradigm has passed four major tests: there are super-horizon perturbations, as shown for the first time in Ref.~\cite{peiris:wmapng}; the power spectrum of these fluctuations is nearly scale invariant~\cite{spergel:wmapcosmoparams} but deviates by a small amount from it, as first shown compellingly in Ref.~\cite{ade:planckcosmoparams2013, ade:planckinflation2013}; the Universe is essentially spatially flat~\cite{hinshaw:wmapcosmoparam, ade:planckcosmoparams2013, ade:planckcosmoparams2015, aghanim:planckcosmoparams2018} and appears homogeneous and isotropic on large scales~\cite{bennett:wmapanomalies, ade:planckanomalies2013, ade:planckanomalies2015}; initial conditions are very nearly Gaussian~\cite{komatsu:testgaussianity, peiris:wmapng, ade:planckng2013, ade:planckng2015, akrami:planckng2018}.

The fact that the inflationary paradigm has passed these tests does not mean it has been verified. Indeed, alternative models exist that also pass the above tests~\cite{steinhardt:cyclicuniverse, brandenberger:stringuniverse}. What is unique of the inflationary paradigm is the existence of an accelerated expansion phase that results in a (quasi) exponential growth of the scale-factor of the metric. This, in turn, facilitates that tensor fluctuations in the metric will manifest themselves as potentially observable gravitational waves~\cite{starobinsky:nosingularity}. This crucial feature of inflation has not yet been measured. Obviously, measuring it would be momentous as it would open up a window into inflation and the early Universe physics not explored before, and would offer the possibility to understand physical mechanisms at play at the energy scale of inflation.

In the simplest inflationary models the amplitude of tensor modes (usually parametrised by the parameter $r$, the tensor-to-scalar ratio at a given scale) can be related to the energy scale of inflation, given by the inflaton potential $V$, by 
\begin{equation}
V^{1/4} = \left(\frac{3}{2} \pi^2 r \mathcal{P}_\zeta\right)^{1/4} M_P \sim 3.3\times 10^{16}\ r^{1/4}\ \mathrm{GeV},
\end{equation}
where $\mathcal{P}_\mathcal{\zeta}$ is the power spectrum\footnote{Here we refer to the almost scale-invariant power spectrum
\begin{equation*}
\mathcal{P}_\zeta = \frac{k^3}{2\pi^2}P_\zeta = \frac{1}{2M_P^2\epsilon} \left(\frac{H_\star}{2\pi}\right)^2 \left(\frac{k}{aH_\star}\right)^{n_s-1},
\end{equation*}
determined by the Hubble expansion rate during inflation $H_\star$ and the slow-roll parameter $\epsilon = \frac{M_P^2}{2}\left(\frac{\partial_\varphi V}{V}\right)^2$, where $\partial_\varphi$ represents the partial derivative with respect to the inflaton field. Past experiments have already measured with great precision the scalar power spectrum amplitude $2\pi^2A_s = \frac{H^2_\star}{4\epsilon M_P^2}$ and the scalar tilt $n_s$. In this work we use $A_s=2.105\cdot 10^{-9}$ and $n_s=0.9665$~\cite{aghanim:planckcosmoparams2018}.} of curvature perturbations on uniform energy density hypersurfaces~$\zeta$ and $M_P = \sqrt{\hbar c/(8\pi G)}$ is the reduced Planck mass. The firm lower limit on the energy scale of inflation is around the MeV scale, to guarantee hydrogen and helium production during Big Bang Nucleosynthesis~\cite{kawasaki:inflationenergyscale, giudice:inflationenergyscale, hannestad:inflationenergyscale, katz:inflationenergyscale}.

An inflationary stochastic background of gravitational waves could in principle be measured directly via future(istic) gravitational wave detection experiments as LISA~\cite{bartolo:inflationwithlisa} (see also~\cite{guzzetti:gwreview, caprini:gwbackground, bartolo:nggwbackground}), DECIGO~\cite{kawamura:decigo} or BBO~\cite{crowder:bbo}, or indirectly via its effect on the polarization of the cosmic microwave background radiation (CMB, see e.g., Ref.~\cite{rees:primevalradiation, kamionkowski:cmbpolarization}). The current observational limit on the tensor-to-scalar ratio is $r\lesssim 0.1$~\cite{ade:planckcosmoparams2015, aghanim:planckcosmoparams2018}. Proposed experiments, as CMBPol~\cite{baumann:cmbpol}, PRISM~\cite{andre:prism} and CORE~\cite{finelli:core}, can reach the $10^{-3}$ level, however it is well known that measuring $r<10^{-4}$ via CMB polarisation is extremely challenging (see e.g., Ref.~\cite{verde:cmbpolarization}) and the cosmic variance limit is at the $10^{-5}$ level~\cite{knox:limitonr}. This implies that the measurement of the CMB polarisation signal can only access inflationary energy scales above $10^{15}$ GeV, only less than an order of magnitude away from the current limit.

A third way one could use to determine the scale of inflation is by probing primordial non-Gaussianities using the information contained in the large-scale structure of the Universe. During the next decade, several galaxy surveys, as DESI~\cite{aghamousa:desisciencebook}, LSST~\cite{abell:lsstsciencebook} and Euclid~\cite{laureijs:euclidsciencebook}, will probe a large volume of our Universe, providing an unprecedented amount of new data. In this context, measuring higher-order statistics, such as the three- or the four-point functions, will extend our knowledge on the inflationary dynamics, which in turn can be used to discriminate between minimal, slow-roll inflationary paradigm and more complex models. On the other hand the specific details of these higher-order statistics can be highly model dependent, therefore the interpretation of the results can be not so straightforward. The non-Gaussian signature arising from  particle exchange  between scalar fluctuations  has recently  received attention  \cite{arkanihamed:particleexchangeI, arkanihamed:particleexchangeII}. In this work we concentrate on a particular non-Gaussian signal called {\it graviton exchange} (GE)~\cite{seery:gravitonexchange}. This signal arises from correlations between inflaton fluctuations mediated by a graviton and enters in the four-point function of scalar curvature perturbations. The magnitude of this non-Gaussian effect is directly proportional to the tensor-to-scalar ratio $r$, therefore by isolating this contribution we can extract a direct information (or a stronger upper bound) on the energy scale of inflation. Moreover, this GE contribution contains much more information about inflationary dynamics, in particular on whether inflation is a strong isotropic attractor, as discussed in Ref. \cite{bordin:tensorconsistencyrelations}.

The paper is organised as follows: in section~\ref{sec:nongaussianity} we review the main results on non-Gaussianities relevant for this work, in section~\ref{sec:dark_matter_halos} we review the framework of excursion regions and halo $n$-points functions and in section~\ref{sec:ge_in_lss} we investigate the magnitude of graviton exchange contribution in large scale structure, in particular to the halo power spectrum~\ref{subsec:halo_power_spectrum} and to the halo bispectrum~\ref{subsec:halo_bispectrum}. Finally we conclude in section~\ref{sec:conclusions}. In section~\ref{app:bispectrum_templates} we discuss bispetrum templates. In this work we use the~$M_P=1$ convention.


\section{Non-Gaussianity}
\label{sec:nongaussianity}
Primordial fluctuations have been found to be consistent with being Gaussian to a very stringent level~\cite{ade:planckng2015, akrami:planckng2018}, however some small deviations from Gaussianity are unavoidable, even in the simplest models, due to the coupling of the inflaton to gravity~\cite{salopek:ngfrominflation, gangui:ngfrominflation, bartolo:ngfrominflation, acquaviva:ngfrominflation, maldacena:ngfrominflation}. The information on how these deviations are created is encoded in the connected part of $n$-point correlators~$\left\langle\zeta_{\mathbf{k}_1}\cdots\zeta_{\mathbf{k}_n}\right\rangle$ (with $n\geq 2$), where $\zeta$ is the curvature perturbation (on uniform energy density hypersurfaces), which is conserved on super-horizon scales for single-field models of inflation. Since curvature perturbations are small (typically $\zeta\sim\mathcal{O}(10^{-5})$ at cosmological scales), it is naively believed that the $(n+1)$-point function is just a small correction to the $n$-point function, however this statement does not take into account the numerous possible mechanisms that can generate a non-Gaussian signal. Moreover, existing small non-Gaussianities can be boosted in the clustering of high density regions that underwent gravitational collapse, as the peaks of the matter density field, that today host virialized structures. 

Since the goal of this work is to provide a new way to constrain the energy scale of inflation, we want to identify some non-Gaussian signal whose strength is directly proportional to the tensor-to-scalar ratio $r$. In particular, in this work we consider the GE contribution to the four-point function and its contribution to the two- and three-point correlation function of collapsed structures. This signal is contaminated by other non-Gaussian signals, such as those coming from the primordial three-point function, which has not been measured yet. For this reason we consider different scenarios, to cover as many inflationary single-field models as possible.

The curvature perturbation $\zeta$, generated by scalar field(s) during inflation, can be connected to the scalar field(s) fluctuation $\delta\varphi$ on an initial spatially-flat hypersurface. The computation of higher-order correlators can be performed using the so-called in-in or Schwinger-Keldysh formalism~\cite{lyth:deltaphiexpansion, schwinger:inin, keldysh:inin, calzetta:inin}, which allows to follow the evolution of the correlators from sub- to super-horizon scales. One can also use other methods, such as second- and higher-order perturbation theory \cite{acquaviva:ngfrominflation, seery:perturbationtheory}, or using the so-called $\delta N$ formalism~\cite{starobinsky:deltan, salopek:ngfrominflation, sasaki:deltan1, sasaki:deltan2, lyth:deltan1, lyth:deltan2}. The latter is equivalent to integrating the evolution of the curvature perturbation on super-horizon scales from horizon exit until some later time after inflation. The correlators of the scalar field(s) fluctuation $\delta\varphi$ at horizon-crossing can then be calculated in an expanding or curved background spacetime using the in-in method. Numerous results have been obtained in this context using these well-established formalisms, both at the level of the bispectrum in single-~\cite{salopek:ngfrominflation, gangui:ngfrominflation, acquaviva:ngfrominflation, maldacena:ngfrominflation, seery:ngsinglefield} and multi-field inflation, see, e.g.,~\cite{bartolo:ngmultifield, seery:ngmultifield, allen:ngmultifield, vernizzi:ngmultifield}, and at the level of the trispectrum in single- and multi-fields inflationary scenarios~\cite{seery:ngtrispectrum, byrnes:ngtrispectrum, arroya:ngtrispectrum}. In this work we consider for simplicity single-field slow-roll inflationary models.

When considering the three-point function, we commonly express it in terms of the bispectrum as
\begin{equation}
\left\langle\zeta_{\mathbf{k}_1}\zeta_{\mathbf{k}_2}\zeta_{\mathbf{k}_3}\right\rangle = (2\pi)^3\delta^D\left(\mathbf{k}_{123}\right)B_\zeta(\mathbf{k}_1,\mathbf{k}_2,\mathbf{k}_3),
\end{equation}
where $\delta^D$ is the Dirac delta, $\mathbf{k}_{ij\dots n}=\mathbf{k}_{i}+\mathbf{k}_{j}+\cdots+\mathbf{k}_{n}$ and the details and the assumptions on the inflationary dynamics are encoded in the $B_\zeta$ function. For completeness, following Ref.~\cite{byrnes:ngtrispectrum}, we also report the curvature bispectrum:
\begin{equation}
B_\zeta(\mathbf{k}_1,\mathbf{k}_2,\mathbf{k}_3) = \left(\partial_\varphi N\right)^3 B_{\delta\varphi}(\mathbf{k}_1,\mathbf{k}_2,\mathbf{k}_3) + (\partial^2_\varphi N)\left(\partial_\varphi N\right)^2\left[P_{\delta\varphi}(k_1)P_{\delta\varphi}(k_2) + (2\ \mathrm{perms.})\right],
\label{eq:complete_bispectrum}
\end{equation}
where $P_{\delta\varphi}$ and $B_{\delta\varphi}$ are the scalar field fluctuation power spectrum and bispectrum, $N$ is the number of e-foldings, $\partial^n_\varphi N\sim \mathcal{O}(\epsilon^{(n-2)/2})$ is the $n$-th derivative of the number of e-folding with respect to the scalar field and it scales with the slow-roll parameter $\epsilon = \frac{1}{2}\left(\partial_\varphi V / V\right)^2$ as indicated. In particular, it has been calculated by Maldacena~\cite{maldacena:ngfrominflation} that in the simplest single-field slow-roll inflationary scenario, at leading order in the slow-roll parameters, the bispectrum reads as
\begin{equation}
B_\zeta^\mathrm{Maldacena}(\mathbf{k}_1,\mathbf{k}_2,\mathbf{k}_3) = \frac{1}{2}\left(\frac{H^2_\star}{4\epsilon}\right)^2\frac{\sum k^3_j}{\prod k^3_j} \left[(1-n_s) + \epsilon\left(\frac{\sum_{i\neq j}k_ik_j^2 + 8\frac{\sum_{i>j}k_i^2k_j^2}{k_t}}{\sum k^3_j} -3 \right) \right]
\label{eq:maldacena_bispectrum}
\end{equation}
where $k_t=\sum_{j=1}^3 k_j$. The term in squared parenthesis, as expected~\cite{bartolo:ngreview}, has a shape-dependent part explicitly suppressed by the slow-roll parameter $\epsilon$. In the limit of one momentum going to zero (squeezed triangular configurations) the term in round parenthesis goes to zero and the whole bispectrum is proportional to~$(1-n_s)$, while in equilateral triangular configurations the same term is maximal and equal to~$5/3$. Typically the entire squared parenthesis is written in terms of a $f^\zeta_\mathrm{NL}$ constant parameter (modulus some proportionality constant), to compare data with theory in a simpler way\footnote{Notice that in the literature there are a series of equivalent, but slightly different parameters. If we would have written the correlators in term of the curvature perturbations on comoving hypersurfaces $\mathcal{R}$ we would have worked with $f^\mathcal{R}_\mathrm{NL}$, while if we have used with the Bardeen's gauge invariant potential $\Phi$, corresponding to the gravitational potential on subhorizon scales, therefore more suitable to work in relation to late times large scale structures, we would have found some constant $f^\Phi_\mathrm{NL}$. Since the three perturbations mentioned above are connected to each other at superhorizon scales by $\Phi = \frac{3(1+w)}{5+3w}\mathcal{R} = -\frac{3(1+w)}{5+3w}\zeta$, the parameters are also connected to each other by $f^\Phi_\mathrm{NL}=f^\mathcal{R}_\mathrm{NL}=-f^\zeta_\mathrm{NL}$, for perturbations entering the horizon during matter domination (if ones uses $\Phi = \Phi_G + f^\Phi_\mathrm{NL} \left[\Phi_G^2 - \left\langle\Phi^2_G\right\rangle \right]$).}. Notice that non-Gaussianities of this type include also a prominent local contribution (the one proportional to~$(1-n_s)$) associated in real space to the well-known quadratic local model~\cite{gangui:ngfrominflation, verde:ngfrominflation, komatsu:ngfrominflation}
\begin{equation}
\zeta = \zeta_G + \frac{3}{5}f^\zeta_\mathrm{NL} \left[\zeta_G^2 - \left\langle\zeta^2_G\right\rangle \right],
\label{eq:quadratic_local_model}
\end{equation}
where $\zeta_G$ is a Gaussian curvature perturbation.

There is a current debate in the literature about whether the $(1-n_s)$ term in equation~\eqref{eq:maldacena_bispectrum} represents the minimum amount of non-Gaussianities that can be observed in the squeezed limit. While some authors argue that it is indeed an intrinsic property of the inflaton that gets imprinted in the dark matter density field~\cite{bartolo:fnlinlss}, others argue that it is simply a gauge quantity that will only manifest itself on higher-order terms with a suppressed value of $f^\zeta_\mathrm{NL} \propto  \left(\frac{k_L}{k_S}\right)^2(1-n_s)$, where $k_L$ and $k_S$ are a long and a short mode, respectively (see e.g., Ref.~\cite{cabass:nongaussianity} and Refs. therein). We point out  that it is still an open question which one is the truly gauge invariant quantity in which the calculation can be performed. It should describe the perturbations behaviour on super-horizon scales and connect the fluctuations in early and late Universe to be used to model the corresponding observables. We also refer the interested reader to Ref.~\cite{abolhasani:nongaussianity}, where a third view on the subject has been presented.

On the other hand, in this work we are mainly interested in the four-point function or trispectrum, in particular its connected part (the disconnected part is always present even in the purely Gaussian case). The complete form of the curvature perturbation trispectrum in single-field inflation, up to second order in slow-roll parameters, reads as~\cite{byrnes:ngtrispectrum}
\begin{equation}
\begin{aligned}
T_\zeta(\mathbf{k}_1,\mathbf{k}_2,\mathbf{k}_3,\mathbf{k}_4) &= (\partial_\varphi N)^4 T_{\delta\varphi}(\mathbf{k}_1,\mathbf{k}_2,\mathbf{k}_3,\mathbf{k}_4)	\\
&+ (\partial^2_\varphi N)(\partial_\varphi N)^3 \left[P_{\delta\varphi}(k_1)B_{\delta\varphi}(k_{12},k_3,k_4) + (11\ \mathrm{perms})\right]	\\
&+ (\partial^2_\varphi N)^2(\partial_\varphi N)^2 \left[P_{\delta\varphi}(k_{13})P_{\delta\varphi}(k_{3})P_{\delta\varphi}(k_{4}) + (11\ \mathrm{perms})\right]	\\
&+ (\partial^3_\varphi N)(\partial_\varphi N)^3 \left[P_{\delta\varphi}(k_{2})P_{\delta\varphi}(k_{3})P_{\delta\varphi}(k_{4}) + (3\ \mathrm{perms})\right],
\end{aligned}
\label{eq:complete_trispectrum}
\end{equation} 
where $T_{\delta\varphi}$ is the scalar field fluctuation trispectrum. By using the linear relation $\zeta\propto \epsilon^{-1/2}\delta\varphi$ we notice that the third and fourth lines of the RHS of equation \eqref{eq:complete_trispectrum} are order $\epsilon^2$ while the order of the first and second line remains to be determine through an explicit computation. The last two lines have also the typical scale dependence coming from the cubic local model in real space:
\begin{equation}
\zeta = \zeta_G + \frac{1}{2}\left(\tau^\zeta_\mathrm{NL}\right)^{1/2} \left[\zeta_G^2 - \left\langle\zeta^2_G\right\rangle \right] + \frac{9}{25}g^\zeta_\mathrm{NL}\left[\zeta_G^3 - 3\zeta_G\left\langle\zeta^2_G\right\rangle \right],
\label{eq:cubic_local_model}
\end{equation}
where we have introduced two non-linearity parameters $\tau^\zeta_\mathrm{NL}$ and $g^\zeta_\mathrm{NL}$ that generate the third and fourth line of equation \eqref{eq:complete_trispectrum}, respectively. These two parameters are expected to be of second order in slow-roll parameters. Finally, notice that only in single-field inflation there is a one-to-one correspondence between $f^\zeta_\mathrm{NL}$ and $\tau^\zeta_\mathrm{NL}$.

In Ref.~\cite{seery:ngtrispectrum} it was demonstrated that the scalar field and the metric remain coupled even in an exact de Sitter space, therefore curvature fluctuations are unavoidably non-Gaussian and there is always a connected four-point function, while naively one would have expected it to be zero. This four-point function is associated to so-called contact interactions, that in terms of Feynman diagrams are associated to a diagram with four scalar external legs. The strength of contact interactions has been roughly estimated to be order $\epsilon$~\cite{seery:ngtrispectrum}, disfavouring the possibility of a detection, however, in successive works~\cite{seery:gravitonexchange, arroya:ngtrispectrum} it has been noticed that nonlinear interactions mediated by tensor fluctuations should also be accounted for, in particular the amplitude of the trispectrum generated by the GE is in general comparable to that generated by contact interactions. More details on the GE contribution can be found in section~\ref{sec:ge_in_lss}.


\section{Dark Matter Halos}
\label{sec:dark_matter_halos}
Even if some level of non-Gaussianity is imprinted in the primordial field $\zeta$, the most relevant quantity for observations is the late-time (smoothed) matter density field. In particular, the effect of non-Gaussianity is enhanced on higher-order correlations of excursion regions which are traced by potentially observable objects such as dark matter halos (or the galaxies these halos host). We define the smoothed linear overdensity field as 
\begin{equation}
\delta_R(\mathbf{x}) = \int d^3y W_R(\mathbf{x}-\mathbf{y})\delta(\mathbf{y}),
\end{equation}
where $W_R$ is a window function of characteristic radius $R$ and $\delta$ is the linear overdensity field. We identify regions corresponding to collapsed objects as those where the smoothed density field exceeds a suitable 
threshold, namely when
\begin{equation}
\delta_R({\bf x}) > \delta_c(z_f) = \frac{\Delta_c(z_f)}{D(z_f)},
\end{equation}
where $z_f$ is the formation redshift of the dark matter halo and we assume that it is very similar to the observed redshift ($z_f\simeq  z_o = z$), $\delta_c(z)$ is the collapse threshold, $\Delta_c(z)$ is the linearly extrapolated overdensity for spherical collapse ($1.686$ in the Einstein-de Sitter and slightly redshift-dependent for more general cosmologies) and $D(z)$ the linear growth factor. The Fourier transform of the (smoothed) linear overdensity field is related to the Bardeen potential $\Phi$ and to the curvature perturbation $\zeta$ via the Poisson equation
\begin{equation}
\delta_{R}(\mathbf{k},z) = \frac{2}{3} \frac{T(k)k^2D(z)}{H_0^2\Omega_{m0}} W_R(k) \Phi(\mathbf{k}) = -\frac{2}{5} \frac{T(k)k^2D(z)}{H_0^2\Omega_{m0}} W_R(k) \zeta(\mathbf{k}) \equiv \mathcal{M}_R(k,z)\zeta(\mathbf{k}),
\end{equation}
where $H_0$ is today's Hubble expansion rate, $\Omega_{m0}$ is the present day matter density fraction, $T(k)$ is the matter transfer function\footnote{In this work we use for the transfer function the analytical estimation provided in Ref.~\cite{eisenstein:transferfunction}, after checking that it does not differ more than $10\%$ at large $k$ from the transfer function obtained from Boltzmann codes as \texttt{CLASS}~\cite{blas:class}. To compute the transfer function we use the cosmological parameters $\omega_\mathrm{b}=0.02242$, $\omega_\mathrm{cdm}~=~0.11933$ and $h=0.6766$~\cite{aghanim:planckcosmoparams2018}.} and $W_R(k)$ is the Fourier transform of the window function in real space~$W_R(r)$\footnote{In this work we use a top-hat filter of radius $R$, of enclosed mass (possibly corresponding to a collapsed object at late times) given by
\begin{equation*}
M = \frac{3H_0^2\Omega_{m0}}{8\pi G}\times\frac{4}{3}\pi R^3.
\end{equation*}
In the rest of this work we use $R=1.824\ \mathrm{Mpc}$, corresponding to $M_\mathrm{halo}=10^{12}\ M_\odot$ dark matter halos. At redshift $z=0$ these halos cannot be considered very massive, however, as we explain in the following section, our goal is to use the information coming from the high redshift Universe, where e.g., $M_\mathrm{halo}=10^{14}\ M_\odot$ dark matter halos (corresponding to $R=8.45\ \mathrm{Mpc}$) are not common. Nevertheless we explicitly checked that at large scales the choice of a different smoothing radius does not change significantly the results.}. The linear growth factor $D(z)$ depends on the background cosmology and reads as $D(z) = (1+z)^{-1}g(z)/g(0)$, where $g(z)$ is the growth suppression factor for non Einstein-de Sitter universes.

The two-point function of the smoothed matter field reads as
\begin{equation}
\left\langle \delta_R(\mathbf{k},z)\delta_R(\mathbf{k}',z)\right\rangle = (2\pi)^3 \delta^D(\mathbf{k}+\mathbf{k}') P_R(k,z),
\end{equation}
where $P_R(k,z) = \mathcal{M}^2_R(k,z) P_{\zeta}(k)$ is the smoothed matter field power spectrum and it is the Fourier transform of the two-point correlation function of the smoothed overdensity field~$\xi_R(r,z)$. Finally, we define the variance of the underlying smoothed overdensity field as 
\begin{equation}
\xi_R(0,z) = \sigma^2_R(z) = \int \frac{d^3k}{(2\pi)^3} P_{R}(k,z).
\end{equation}

For Gaussian or slightly non-Gaussian fields, virtually all regions above a high threshold are peaks and therefore will eventually host virialized structures (i.e., massive dark matter halos). Non-Gaussianities change the clustering properties of halos. For regions above a high threshold (and therefore to an extremely good approximation for massive halos), the two-point correlation function reads~\cite{grinstein:ngcorrelation, matarrese:ngcorrelation, lucchin:ngcorrelation}
\begin{equation}
\xi_\mathrm{halo}(\mathbf{r}) = \exp\left[ \sum_{N=2}^{\infty} \sum_{j=1}^{N-1} \frac{\nu^N\sigma_R^{-N}}{j!(N-j)!} \xi_R^{(N)} (\underbrace{\mathbf{x}_1, ..., \mathbf{x}_1}_{j\, \mathrm{times}}, \underbrace{\mathbf{x}_2, ..., \mathbf{x}_2}_{(N-j)\, \mathrm{times}}) \right] - 1,
\label{eq:xi_ng}
\end{equation}
where $\mathbf{r} = \mathbf{x}_1 - \mathbf{x}_2$, $\nu(z,M) = \Delta_c(z) / \sigma_R(z)$ is the dimensionless peak height, $\xi_R^{(N)}~=~\langle\underbrace{\delta_R\cdots\delta_R}_{N\ \mathrm{times}}\rangle$ are the $N$-point connected correlation functions and $\xi_R^{(2)}\equiv\xi_R$. The generalization of equation~\eqref{eq:xi_ng} to the three-point correlation function is~\cite{matarrese:ngcorrelation} 
\begin{equation}
\begin{aligned}
\Xi_\mathrm{halo}(\mathbf{x}_1,\mathbf{x}_2,\mathbf{x}_3) &= F(\mathbf{x}_1,\mathbf{x}_2,\mathbf{x}_3) \left[\prod_{i<j}\xi_\mathrm{halo}(\mathbf{x}_i,\mathbf{x}_j) + \left[ \xi_\mathrm{halo}(\mathbf{x}_1,\mathbf{x}_2) \xi_\mathrm{halo}(\mathbf{x}_2,\mathbf{x}_3) + (2\ \mathrm{perms.}) \right]\right] \\
&\qquad+ \left[F(\mathbf{x}_1,\mathbf{x}_2,\mathbf{x}_3) - 1\right]\left[\sum_{i<j}\xi_\mathrm{halo}(\mathbf{x}_i,\mathbf{x}_j) + 1\right],
\end{aligned}
\end{equation}
where
\begin{equation}
F(\mathbf{x}_1,\mathbf{x}_2,\mathbf{x}_3) = \exp\left[\sum_{N=3}^{\infty} \sum_{j=1}^{N-2} \sum_{k=1}^{N-j-1} \frac{\nu^N\sigma_R^{-N}}{j!k!(N-j-k)!} \xi_R^{(N)} (\underbrace{\mathbf{x}_1, ..., \mathbf{x}_1}_{j\, \mathrm{times}}, \underbrace{\mathbf{x}_2, ..., \mathbf{x}_2}_{k\, \mathrm{times}}, \underbrace{\mathbf{x}_3,\ ...,\ \mathbf{x}_3}_{(N-j-k)\, \mathrm{times}}) \right].
\label{eq:F_form}
\end{equation}
Here we notice that the $N$-th order term scales with redshift as $D(z)^{-N}$, hence going to high redshift we observe enhanced non-Gaussian features with respect to redshift $z=0$. In fact, from our definitions, we have that $(\nu/\sigma_R)^N\propto D(z)^{-2N}$ and $\xi^{(N)}_R\propto D(z)^{N}$, since in the $N$-point function each $\delta_R$ comes along with a $D(z)$ factor, independently on the Gaussian or non-Gaussian origin of such $N$-point connected correlation function. Therefore going to higher redshift boosts the non-Gaussian signal with respect to its magnitude at redshift $z=0$, even if we don't expand the exponential in equations \eqref{eq:xi_ng} and \eqref{eq:F_form}.

In the limit of purely Gaussian initial conditions, where $\xi_R^{(N\geq 3)}\equiv 0$ hence $F(\mathbf{x}_1,\mathbf{x}_2,\mathbf{x}_3)=1$, the two-~\cite{kaiser:correlation, politzer:correlation, jensen:correlation} and three-point~\cite{matarrese:ngcorrelation} functions of excursion regions becomes 
\begin{equation}
\begin{aligned}
\xi^{G}_\mathrm{halo}(\mathbf{r}) &= \exp\left[\frac{\nu^2}{\sigma_R^2}\xi^{(2)}_R(\mathbf{r})\right] - 1,	\\
\Xi^{G}_\mathrm{halo}(\mathbf{x}_1,\mathbf{x}_2,\mathbf{x}_3) &= \left[\prod_{i<j}\xi^{G}_\mathrm{halo}(\mathbf{x}_i,\mathbf{x}_j) + \left[ \xi^{G}_\mathrm{halo}(\mathbf{x}_1,\mathbf{x}_2) \xi^{G}_\mathrm{halo}(\mathbf{x}_2,\mathbf{x}_3) + (2\ \mathrm{perms.}) \right]\right].
\end{aligned}
\label{eq:gaussian_23PF}
\end{equation}
The above equations are typically expanded in the limit of high-density peaks $(\nu\gg 1)$ and large separation between halos (large scale limit, $r\gg R$, where $\xi^{(N)}_R\ll 1$). In this limit, we expect $\delta_R$ to be small, therefore we can identify it as a small parameter in which the expansion is done and we can roughly estimate the $N$-point correlation functions as $\xi_R^{(N)}\sim \mathcal{O}(\delta_R^N)$. We choose to expand equations~\eqref{eq:gaussian_23PF} up to second order, to check that higher order corrections do not contaminate the non-Gaussian signal we are interested in. In particular for the two- and three-point point correlation functions we obtain
\begin{equation}
\begin{aligned}
\xi^{G}_\mathrm{halo}(\mathbf{r}) &\approx b^2_L\xi^{(2)}_R(\mathbf{r}) + \frac{b_L^4}{2}\left[\xi^{(2)}_R(\mathbf{r})\right]^2,	\\
\Xi^{G}_\mathrm{halo}(\mathbf{x}_1,\mathbf{x}_2,\mathbf{x}_3) &\approx b^4_L\left[ \xi^{(2)}_R(\mathbf{x}_1,\mathbf{x}_2) \xi^{(2)}_R(\mathbf{x}_2,\mathbf{x}_3) + (2\ \mathrm{perms.}) \right]	\\
&\quad + b^6_L \xi^{(2)}_R(\mathbf{x}_1,\mathbf{x}_2) \xi^{(2)}_R(\mathbf{x}_2,\mathbf{x}_3) \xi^{(2)}_R(\mathbf{x}_1,\mathbf{x}_3)	\\
&\quad +\frac{b_L^6}{2}\left[ \xi^{(2)}_R(\mathbf{x}_1,\mathbf{x}_2) \left[\xi^{(2)}_R(\mathbf{x}_2,\mathbf{x}_3)\right]^2 + (2\ \mathrm{perms.}) \right],
\end{aligned}
\label{eq:gaussian_34PF}
\end{equation}
where $b_{L}(z) = \nu(z)/\sigma_R(z) = \Delta_c(z)/\sigma^2_R(z)$ is the Lagrangian linear bias. As noted for the first time by the authors of Ref.~\cite{matarrese:ngcorrelation}, even if initial conditions are perfectly Gaussian, the three-point correlation function of excursion regions is non-zero and constitutes an unavoidable background signal from which the true primordial non-Gaussian signal has to be extracted. We further analyse the form of the Gaussian part in section~\ref{subsec:halo_bispectrum}, however we stress that it is not unexpected for the filtering procedure to introduce some feature in correlations functions of all orders, since the smoothing procedure is highly nonlocal and nonlinear. We refer the interested reader to Ref.~\cite{verde:biasweightedhalos}, where the authors investigate the effects of the smoothing procedure on dark matter halos bias. 

On the other hand, for non-Gaussian initial conditions, other terms appear in the above Taylor expansion. By expanding up to $N=4$ order to include the four-point correlation function contribution, we have that the non-Gaussian part of the two- and three-point functions read as \cite{jeong:oneloopcorrections, sefusatti:oneloopcorrections}
\begin{equation}
\begin{aligned}
\xi^{NG}_\mathrm{halo}(\mathbf{r}) &\approx \xi^{G}_\mathrm{halo}(\mathbf{r}) + b_L^3 \xi_R^{(3)}(\mathbf{x}_1, \mathbf{x}_1, \mathbf{x}_2)	+ b_L^4\left[\frac{\xi^{(4)}_R(\mathbf{x}_1, \mathbf{x}_1, \mathbf{x}_1, \mathbf{x}_2)}{3} + \frac{\xi^{(4)}_R(\mathbf{x}_1, \mathbf{x}_1, \mathbf{x}_2, \mathbf{x}_2)}{4}\right]	\\
&\quad + b_L^5 \xi_R^{(2)}(\mathbf{x}_1, \mathbf{x}_2)\xi_R^{(3)}(\mathbf{x}_1, \mathbf{x}_1, \mathbf{x}_2),
\end{aligned}
\label{eq:halo_twopoint_correlation}
\end{equation}
\begin{equation}
\begin{aligned}
\Xi^{NG}_\mathrm{halo}(\mathbf{x}_1,\mathbf{x}_2,\mathbf{x}_3) &\approx \Xi^{G}_\mathrm{halo}(\mathbf{x}_1,\mathbf{x}_2,\mathbf{x}_3) + b_L^3\xi^{(3)}_R(\mathbf{x}_1,\mathbf{x}_2,\mathbf{x}_3) \\
&\quad + b_L^4\left[\frac{\xi^{(4)}_R(\mathbf{x}_1,\mathbf{x}_1,\mathbf{x}_2,\mathbf{x}_3)}{2} + \frac{\xi^{(4)}_R(\mathbf{x}_1,\mathbf{x}_2,\mathbf{x}_2,\mathbf{x}_3)}{2} + \frac{\xi^{(4)}_R(\mathbf{x}_1,\mathbf{x}_2,\mathbf{x}_3,\mathbf{x}_3)}{2}\right]	\\
&\quad + b_L^5\xi^{(3)}_R(\mathbf{x}_1,\mathbf{x}_2,\mathbf{x}_3)\sum_{i<j}\xi^{(2)}_R(\mathbf{x}_i,\mathbf{x}_j) ,
\end{aligned}
\label{eq:halo_threepoint_correlation}
\end{equation}
where in the last lines of equations \eqref{eq:halo_twopoint_correlation} and \eqref{eq:halo_threepoint_correlation} we report also the first Gaussian/non-Gaussian mixed contribution, even if it is expected to be one order of magnitude lower in~$\delta_R$ than the trispectrum contribution. To leading order, the non-Gaussian correction to the $n$-points functions of massive halos is a -truncated- sum of contributions of the three- and four-point (primordial) functions, enhanced by powers (third and fourth powers respectively) of bias,~$b_L$. Notice that so far these results are very generic, in fact the equations above do not assume any specific origin of the three- and four-point correlation functions and constitute the starting point of our analysis.

The validity of the approach described above has been repeatedly tested against numerical simulations with Gaussian and non-Gaussian initial conditions, finding that theory agrees with simulations. In particular, on large enough scales, we have that $b_L^N\xi^{(N)}_R$ is small and the series expansion does not have convergence issues. The interested reader can check e.g., Refs.~\cite{dalal:simulations, desjacques:simulations, grossi:simulations, pillepich:simulations, nishimichi:simulations, wagner:bispectrumtemplatesI, wagner:bispectrumtemplatesII}.


\section{Graviton Exchange Signal in Large Scale Structure}
\label{sec:ge_in_lss}
The trispectrum generated by GE was derived in detail in Ref.~\cite{seery:gravitonexchange}. Very recently Baumann and collaborators \cite{arkanihamed:particleexchangeII} re-derived the  GE-induced higher order correlations in a more general context. We leave the analysis of their findings to future work and consider here the GE trispectrum of Ref.~\cite{seery:gravitonexchange}. In principle there are two distinct ways to measure the GE contribution in large scale structure data. 

\begin{figure}[h!]
\centerline{
\subfloat{
	\begin{tikzpicture}
		\begin{feynman}[large]
		\vertex (a);
		\vertex [above=1.5cm of a] (i1);
		\vertex [right=2.5cm of i1] (b);
		\vertex [below=1.5cm of a] (i2);
		\vertex [right=3.5cm of i2] (c);
		\vertex [right=7.0cm of a] (d);
		\diagram* {
		(a) -- [fermion, edge label=\(\mathbf{k}_2\)] (b) -- [fermion, edge label=\(\mathbf{k}_3\)] (d),
		(a) -- [anti fermion, edge label'=\(\mathbf{k}_1\)] (c) -- [anti fermion, edge label'=\(\mathbf{k}_4\)] (d),
		};
		\end{feynman}
	\end{tikzpicture}}
\subfloat{
	\begin{tikzpicture}
		\begin{feynman}[large]
		\vertex (a);
		\vertex [above=0.25cm of a] (i1);
		\vertex [right=2.5cm of i1] (b);
		\vertex [below=0.25cm of a] (i2);
		\vertex [right=2.9cm of i2] (c);
		\vertex [right=7.0cm of a] (d);
		\diagram* {
		(a) -- [fermion, edge label=\(\mathbf{k}_2\)] (b) -- [fermion, edge label=\(\mathbf{k}_3\)] (d),
		(a) -- [anti fermion, edge label'=\(\mathbf{k}_1\)] (c) -- [anti fermion, edge label'=\(\mathbf{k}_4\)] (d),
		};
		\end{feynman}
	\end{tikzpicture}}}
\caption{Kite (\textit{left panel}) and folded kite (\textit{right panel}) diagrams. In the left diagram we have $k_{13}\ll k_1\sim k_3, k_2 \sim k_4$, while in the right one we have $k_{12}\ll k_1\sim k_2, k_3 \sim k_4$. Diagrams have been drawn with Ti$k$Z-Feynman \cite{ellis:tikzfeynman}.}
\label{fig:trispectrum_configurations}
\end{figure}
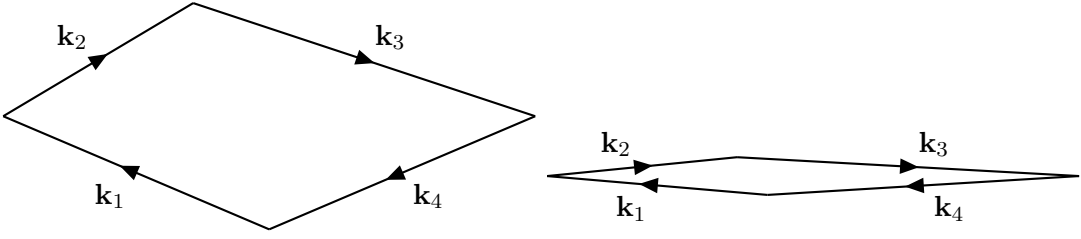

The first one is to look for it directly in the trispectrum of the dark matter or low-to-moderate biased tracers of it. In this case, as pointed out by Ref.~\cite{seery:gravitonexchange} some configurations are particularly interesting and well suited since the size of non-Gaussianity is amplified. These configurations are associated to the so-called counter-collinear limit, where the sum of two momenta goes to zero (e.g., when $k_{12}\ll k_1\approx k_2, k_3\approx k_4$). We show in figure~\ref{fig:trispectrum_configurations} the two possible (dual) configurations, called \textit{kite}, if the momenta summing up to zero are on opposite sides of the parallelogram, and \textit{folded kite}, if the momenta summing up to zero are on contiguous side of the parallelogram. In these configurations, where all momenta are finite, the GE contribution diverges (e.g., scaling as $k_{12}^{-3}$) opening the possibility for amplifying the signal.

A direct measurement of the primordial trispectrum has been done at the CMB level in Refs.~\cite{ade:planckng2013, ade:planckng2015, akrami:planckng2018, kunz:cmbtrispectrum, detroia:cmbtrispectrum, smidt:cmbtrispectrum, sekiguchi:cmbtrispectrum}. However doing so from large-scale structure surveys may be challenging because of the number of trispectrum modes involved and the low-signal to noise per mode; for this reason very few attempt have been done so far~\cite{verde:measuringtrispectrum}. 

In this section we consider the alternative approach of looking at the effect of the GE trispectrum contribution in the halo two- and three-point functions. The trispectrum due to a graviton exchange is given by \cite{seery:gravitonexchange}
\begin{equation}
\begin{aligned}
\left\langle \zeta_{\mathbf{k}_1}\zeta_{\mathbf{k}_2}\zeta_{\mathbf{k}_3}\zeta_{\mathbf{k}_4}\right\rangle^\mathrm{GE} &= (2\pi)^3 \delta(\mathbf{k}_{1234}) \left(\frac{H_\star^2}{4\epsilon}\right)^3 \frac{r/4}{\prod_j k_j^3}\times	\\
\times &\left[\frac{k_1^2 k_3^2}{k^3_{12}}\left[1-(\hat{\mathbf{k}}_{1}\cdot\hat{\mathbf{k}}_{12})^2\right] \left[1-(\hat{\mathbf{k}}_{3}\cdot\hat{\mathbf{k}}_{12})^2\right] \cos 2\chi_{12,34} \cdot (\mathcal{I}_{1234}+\mathcal{I}_{3412}) +	\right.	\\
&+ \frac{k_1^2 k_2^2}{k^3_{13}} \left[1-(\hat{\mathbf{k}}_{1}\cdot\hat{\mathbf{k}}_{13})^2\right] \left[1-(\hat{\mathbf{k}}_{2}\cdot\hat{\mathbf{k}}_{13})^2\right]
\cos 2\chi_{13,24} \cdot (\mathcal{I}_{1324}+\mathcal{I}_{2413}) + \\
&\left. + \frac{k_1^2 k_2^2}{k^3_{14}} \left[1-(\hat{\mathbf{k}}_{1}\cdot\hat{\mathbf{k}}_{14})^2\right] \left[1-(\hat{\mathbf{k}}_{2}\cdot\hat{\mathbf{k}}_{14})^2\right]
\cos 2\chi_{14,23} \cdot (\mathcal{I}_{1423}+\mathcal{I}_{2314}) \right] ,
\end{aligned}
\label{eq:graviton_exchange_trispectrum}
\end{equation}
where $\cos\chi_{ij,kl}=(\hat{\mathbf{k}}_i\times \hat{\mathbf{k}}_j) \cdot (\hat{\mathbf{k}}_k\times \hat{\mathbf{k}}_l)$ is the angle between the two planes formed by $\left\lbrace\mathbf{k}_i,\mathbf{k}_j\right\rbrace$ and $\left\lbrace\mathbf{k}_k,\mathbf{k}_l\right\rbrace$,
\begin{equation}
\begin{aligned}
\mathcal{I}_{1234} + \mathcal{I}_{3412} &= \frac{k_1+k_2}{a_{34}^2}\left[\frac{1}{2} (a_{34}+k_{12})(a_{34}^2-2b_{34}) + k_{12}^2 (k_3+k_4) \right] + (1,2 \leftrightarrow 3,4)\\
&+ \frac{k_1k_2}{k_t} \left[\frac{b_{34}}{a_{34}} - k_{12} + \frac{k_{12}}{a_{12}} \left(k_3k_4 -k_{12} \frac{b_{34}}{a_{34}}\right) \left(\frac1{k_t} +\frac1{a_{12}}
\right)\right] + (1,2 \leftrightarrow 3,4) \\
& - \frac{k_{12}}{a_{12} a_{34} k_t} \left[ b_{12} b_{34} + 2k_{12}^2 k_p \left(\frac{1} {k_t^2}+\frac{1}{a_{12} a_{34}} +\frac{k_{12}}{k_t a_{12} a_{34}} \right)\right],
\end{aligned}
\label{eq:integrand_shape}
\end{equation}
$a_{ij} = k_i + k_j + k_{ij}$, $b_{ij} = (k_i + k_j)k_{ij}$, $k_t = \sum_{j=1}^{4} k_j$ and $k_p=\prod_{j=1}^{4} k_j$\footnote{Note that scalar and vector products in the above equation can be uniquely computed using spherical coordinates as
\begin{equation*}
\hat{\mathbf{k}}_i\cdot\hat{\mathbf{k}}_j = \sin\theta_i\sin\theta_j\cos(\phi_i-\phi_j)+\cos\theta_i\cos\theta_j,
\end{equation*}
\begin{equation*}
\begin{aligned}
\hat{\mathbf{k}}_i\times \hat{\mathbf{k}}_j &= \left( \sin\theta_i\sin\phi_i\cos\theta_j - \cos\theta_i\sin\theta_j\sin\phi_j \right)\hat{\mathbf{k}}_x	\\
&\quad +\left( \cos\theta_i\sin\theta_j\cos\phi_j - \sin\theta_i\cos\phi_i\cos\theta_j \right)\hat{\mathbf{k}}_y	\\
&\quad +\ \sin\theta_i\sin\theta_j\sin(\phi_j-\phi_i) \hat{\mathbf{k}}_z.
\end{aligned}
\end{equation*}
}.

In principle there are a multitude of late-time, non-primordial effects that should be taken into account when measuring non-Gaussianity in large scale structure. Here, we are interested in estimating only the size of specific effects, and we refer the interested reader e.g., to Ref.~\cite{karagiannis:nganalysis} for a comprehensive analysis.

In this work we use the public \texttt{Cubature}\footnote{The package has be written by Steven G. Johnson and can be found in GitHub \href{https://github.com/stevengj/cubature}{https://github.com/stevengj/cubature}.} package to compute the multidimensional integrals. Notice that in doing the integrals, besides the obvious singularity when one of the momenta goes to zero that the package can easily deal with, there is another singularity, i.e., the counter-collinear limit, when the sum of two momenta goes to zero. Since the region where this happens has some non-trivial shape, we decided to regularize the integrand close to the singularity by multiplying each term (lines two, three and four) in equation~\eqref{eq:graviton_exchange_trispectrum} by $e^{-k_\mathrm{hor}/k_{ij}}$, where $k_\mathrm{hor}$ is a mode entering the horizon at late time and $k_{ij}$ is the respective momentum at the denominator. The physical interpretation of such regularization is straightforward: we cannot probe wave numbers smaller than those that are crossing the horizon today, since smaller wave numbers appear as an uniform background. In doing so we are removing extremely folded configurations (that might be related to ``gauge-invariance'' considerations). On the one hand our regularisation method artificially suppresses modes $k\lesssim k_\mathrm{hor}$, on the other hand we explicitly checked that this procedure does not introduce any significant bias in the magnitude of the GE contribution when $k\gg k_\mathrm{hor}$. We choose $k_\mathrm{hor}~=~10^{-6}\ \mathrm{Mpc}^{-1}$, much less than $k^\mathrm{today}_\mathrm{hor}\sim \mathcal{O}(10^{-4})\ \mathrm{Mpc}^{-1}$, in order not to affect the modes that are of cosmological interest. This phenomenological procedure represents a first attempt to tackle the long-standing problem of a correct treatment of super-horizon modes. The improvement of this method is left for future work. 

The GE contribution was derived in the context of standard single-field slow-roll inflation, namely using standard kinetic term, no modified gravity, Bunch-Davies vacuum and others~\cite{seery:gravitonexchange}. However, in order to help the readers to compare these contributions to other bispectrum templates they may be familiar with, we include in the figures of the following sections also the bispectrum templates of appendix~\ref{app:bispectrum_templates}, which arise when different assumptions are taken. We choose as reference values for non-Gaussianity parameters $r=0.1$ (maximum value allowed by current CMB data~\cite{aghanim:planckcosmoparams2018}), $\epsilon = r/16 = 0.00625$ and $|f^\zeta_\mathrm{NL}|=(1-n_s)/12=0.00279$. It should be noticed that Cosmic Microwave Background data currently allow higher values of $|f^\zeta_\mathrm{NL}|\sim\mathcal{O}(1-10)$, depending on the bispectrum template, see e.g., Ref.~\cite{ade:planckng2015}. However, in the cases we are interested in, $r$ and $f^\zeta_\mathrm{NL}$ act only as an overall amplitude rescaling factor, therefore the reader can simply shift vertically the lines to match with the desired value of such parameters.


\subsection{Signal in the Halo Power Spectrum}
\label{subsec:halo_power_spectrum}
To compute the halo power spectrum, in the equations below we take the Fourier transform $(\mathrm{FT}\left\lbrace\cdot\right\rbrace)$ of equation~\eqref{eq:halo_twopoint_correlation}, 
\begin{equation}
P^{NG}_\mathrm{halo}(k,z) \approx P^G_\mathrm{halo}(k,z) + B_\mathrm{112}(k,z) + T_\mathrm{1112}(k,z) + T_\mathrm{1122}(k,z) + M_{12-112}(k,z),
\label{eq:nongaussian_halo_powerspectrum}
\end{equation}
where we recognise  the Gaussian halo power spectrum,
\begin{equation}
P^G_\mathrm{halo}(k,z) \approx b_L^2(z) P_R(k,z) + \frac{b_L^4(z)}{2}\int\frac{d^3q}{(2\pi)^3} P_R(q,z)P_R(|\mathbf{k}-\mathbf{q}|,z),
\label{eq:gaussian_halo_powerspectrum}
\end{equation}
the purely non-Gaussian contributions, 
\begin{equation}
\begin{aligned}
 B_\mathrm{112}(k,z) &= b_L^3(z) \mathrm{FT}\left\lbrace\xi_R^{(3)}(\mathbf{x}_1, \mathbf{x}_1, \mathbf{x}_2)\right\rbrace =	\\
&= b_L^3(z) \int \frac{d^3q}{(2\pi)^3} \mathcal{M}_R(q,z)\mathcal{M}_R(|\mathbf{k}-\mathbf{q}|,z)\mathcal{M}_R(k,z)B_\zeta(\mathbf{q},\mathbf{k}-\mathbf{q},-\mathbf{k}),	\\
 T_\mathrm{1112}(k,z) &= \frac{b_L^4(z)}{3} \mathrm{FT}\left\lbrace\xi_R^{(4)}(\mathbf{x}_1, \mathbf{x}_1, \mathbf{x}_1, \mathbf{x}_2)\right\rbrace =	\\
&= \frac{b_L^4(z)}{3} \int \frac{d^3q_1}{(2\pi)^3}\frac{d^3q_2}{(2\pi)^3} \mathcal{M}_R(q_1,z)\mathcal{M}_R(q_2,z)\mathcal{M}_R(|\mathbf{k}-\mathbf{q}_{12}|,z)\mathcal{M}_R(k,z)\times	\\
&\qquad\qquad\qquad\qquad\qquad\qquad \times T_\zeta(\mathbf{q}_{1},\mathbf{q}_{2},\mathbf{k}-\mathbf{q}_{12},-\mathbf{k}),	\\
 T_\mathrm{1122}(k,z) &= \frac{b_L^4(z)}{4} \mathrm{FT}\left\lbrace\xi_R^{(4)}(\mathbf{x}_1, \mathbf{x}_1, \mathbf{x}_2, \mathbf{x}_2)\right\rbrace =	\\
&= \frac{b_L^4(z)}{4} \int \frac{d^3q_1}{(2\pi)^3}\frac{d^3q_2}{(2\pi)^3} \mathcal{M}_R(|\mathbf{k}-\mathbf{q}_1|,z)\mathcal{M}_R(q_1,z)\mathcal{M}_R(q_2,z)\mathcal{M}_R(|\mathbf{k}+\mathbf{q}_2|,z)\times	\\
&\qquad\qquad\qquad\qquad\qquad\qquad \times T_\zeta(\mathbf{k}-\mathbf{q}_1,\mathbf{q}_{1},\mathbf{q}_{2},-\mathbf{k}-\mathbf{q}_2),
\end{aligned}
\end{equation}
and the mixed contribution
\begin{equation}
\begin{aligned}
M_{12-112}(k,z) &= b_L^5(z) \mathrm{FT}\left\lbrace\xi_R^{(2)}(\mathbf{x}_1, \mathbf{x}_2)\xi_R^{(3)}(\mathbf{x}_1, \mathbf{x}_1, \mathbf{x}_2)\right\rbrace \\
&= b_L^5(z) \int \frac{d^3q_1}{(2\pi)^3}\frac{d^3q_2}{(2\pi)^3} P_R(|\mathbf{k}-\mathbf{q}_{12}|,z)B_R(q_1,q_2,q_{12},z).
\end{aligned}
\label{eq:ps_mixed_contribution}
\end{equation}

\begin{figure}[h]
\centering
\includegraphics[width=\textwidth]{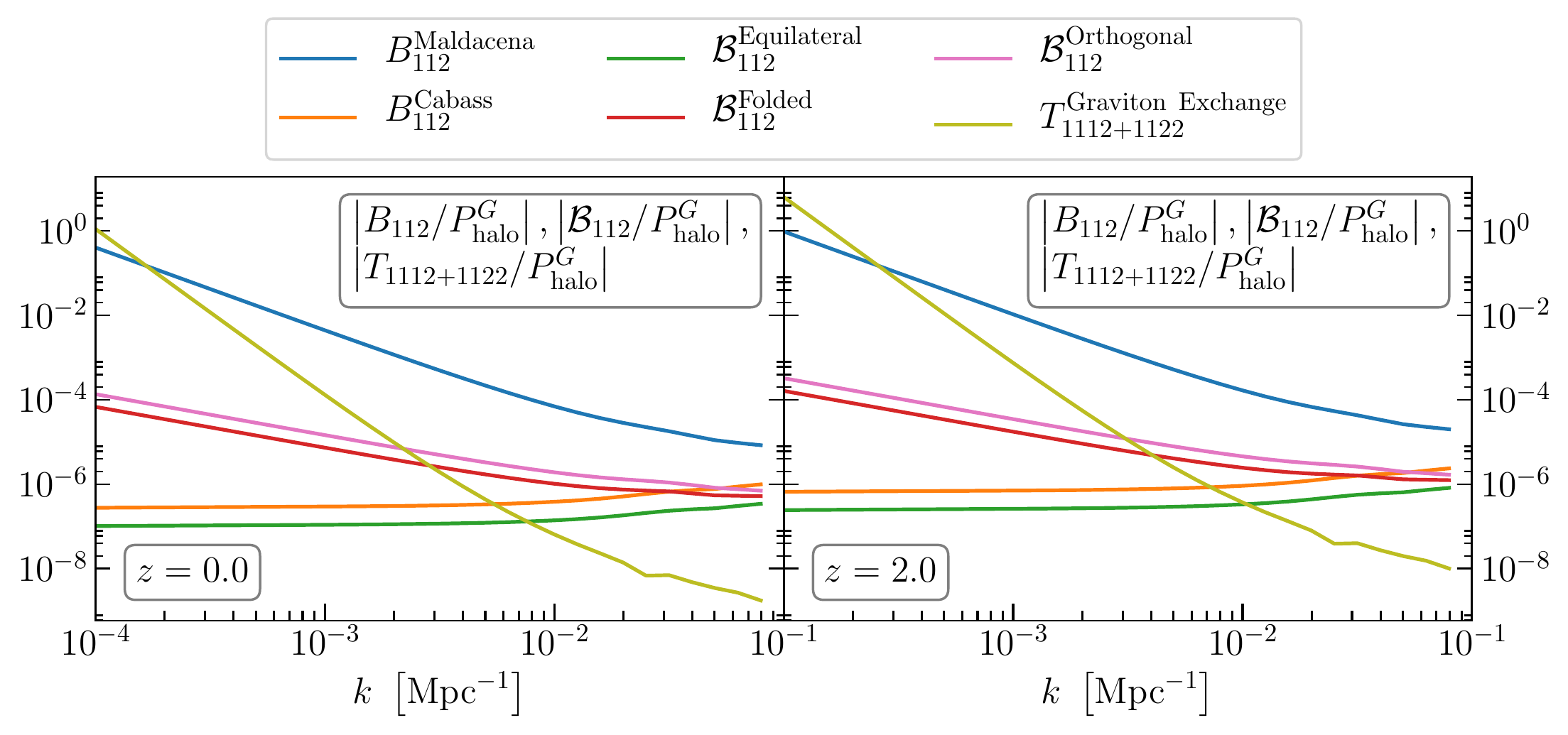}
\caption{Ratio between different primordial non-Gaussian contribution (bispectra and GE trispectrum) and the Gaussian halo power spectrum at redshift $z=0$ (\textit{left panel}) and $z=2$ (\textit{right panel}) for $M_\mathrm{halo}=10^{12}\ M_{\odot}$ dark matter halos. For the Maldacena and Cabass bispectra, indicated by $B_{112}$, we use~$\epsilon~=~0.00625$, while for the Equilateral, Folded and Orthogonal templates, indicated by $\mathcal{B}_{112}$, we use $f_\mathrm{NL}=0.00279$. In the case of the templates, a different value of $f_\mathrm{NL}$ would simply rescale vertically the lines. For the GE contribution we use~$r=0.1$ and $k_\mathrm{hor}=10^{-6}\ \mathrm{Mpc}^{-1}$. Also in this case different values of $r$ simply rescales vertically the GE contribution.}
\label{fig:GE_in_PS}
\end{figure}

In the context of quadratic and cubic models of local non-Gaussianities of equations~\eqref{eq:quadratic_local_model} and~\eqref{eq:cubic_local_model}, the three- and the four-points contribution has already been evaluated by Refs.~\cite{matarrese:nghalobias, desjacques:ngsignature}, respectively. We compute the GE contribution following the same procedure, by substituting equation~\eqref{eq:graviton_exchange_trispectrum} into the four-point correlation function on the RHS of equation~\eqref{eq:nongaussian_halo_powerspectrum}. Since we are interested only in primordial features, we report in figure~\ref{fig:GE_in_PS} the ratio between the primordial non-Gaussian contributions of equation~\eqref{eq:nongaussian_halo_powerspectrum} and the Gaussian halo power spectrum at different redshift, to compare the relative strength of the signals coming from Gaussian and non-Gaussian processes and the relative strength of the bispectrum and trispectrum terms.

Notice that from an operational point of view, the GE signal should be extracted from the total halo power spectrum by subtracting the bispectrum contribution, which in this case acts as an additional source of ``noise''. As we explained in section~\ref{sec:nongaussianity}, there is an ongoing debate in the literature on the correct form of the bispectrum in the local case, therefore we report both possibilities. Following Cabass~\cite{cabass:nongaussianity}, we multiply the $(1-n_s)$ factor in equation~\eqref{eq:maldacena_bispectrum} by an additional factor $(k_\mathrm{longest}/k_\mathrm{shortest})^2$, where $k_\mathrm{longest}$ and $k_\mathrm{shortest}$ are the longest and shortest modes of the considered triangle. We are aware  that the GE contribution has not been computed under  different  assumptions, for example the conditions that give rise to different bispectra shapes such as non Bunch-Davies vacuum states. However  the authors of Ref.~\cite{seery:gravitonexchange} indicate that their results can be extended to more general conditions. Here, for helping the reader to compare these contributions to other bispectra they may be familiar with, we have included also the bispectrum templates, $\mathcal{B}$, defined in equations~\eqref{eq:equilateral_bispectrum},~\eqref{eq:folded_bispectrum} and~\eqref{eq:orthogonal_bispectrum}, which have already been studied in the halo power spectrum context for instance in Refs.~\cite{verde:bispectrummodels, schmidt:bispectrummodels}. As it can be seen in figure~\ref{fig:GE_in_PS}, depending on the specific model and magnitude of primordial non-Gaussianities, the GE contribution is comparable to or even larger than the primordial bispectrum signal at the largest scales. By comparing the two panels, we also notice that the importance of the GE increases with redshift.

Although a detailed signal-to-noise and survey forecast calculation is well beyond the scope of this paper, figure \ref{fig:GE_in_PS} indicates that the GE contribution can be singled out and extracted from the measured halo power spectrum thanks to the different scale dependence of the terms in equation \eqref{eq:nongaussian_halo_powerspectrum}. In particular, at large scales, we have that
\begin{equation}
\begin{aligned}
&B^\mathrm{Cabass}_\mathrm{112}/P^G_\mathrm{halo},\ \mathcal{B}^\mathrm{Equilateral}_\mathrm{112}/P^G_\mathrm{halo} \propto k^{0}(1+z)\frac{g(0)}{g(z)},	\\
&\mathcal{B}^\mathrm{Orthogonal}_\mathrm{112}/P^G_\mathrm{halo},\ \mathcal{B}^\mathrm{Folded}_\mathrm{112}/P^G_\mathrm{halo} \propto k^{-1}(1+z)\frac{g(0)}{g(z)},	\\
&B^\mathrm{Maldacena}_\mathrm{112}/P^G_\mathrm{halo} \propto k^{-2}(1+z)\frac{g(0)}{g(z)},	\\
\end{aligned}
\label{eq:bispectrum_over_powerspectrum_scaling}
\end{equation}
while the two trispectrum contributions scale as
\begin{equation}
\begin{aligned}
T_\mathrm{1112}/P^G_\mathrm{halo} &\propto k^{-2}\left[(1+z)\frac{g(0)}{g(z)}\right]^2,	\\
T_\mathrm{1122}/P^G_\mathrm{halo} &\propto k^{-4}\left[(1+z)\frac{g(0)}{g(z)}\right]^2.
\end{aligned}
\label{eq:trispectrum_over_powerspectrum_scaling}
\end{equation}
We have checked that for all the cases of interest, that is bias of order few, the second term in equation~\eqref{eq:gaussian_halo_powerspectrum} is subdominant with respect to the first one that scales as $D^{-2}$, therefore in equations~\eqref{eq:bispectrum_over_powerspectrum_scaling} and~\eqref{eq:trispectrum_over_powerspectrum_scaling} only the dominant term matters. Notice also that in equation~\eqref{eq:trispectrum_over_powerspectrum_scaling} the term $T_\mathrm{1122}(k)$ dominates over the $T_\mathrm{1112}(k)$ term at large scales and it has a scale dependence different from any other common bispectrum template. Other terms of the trispectrum could have the same scale dependence, e.g., the terms in the third line of equation~\eqref{eq:complete_trispectrum}, as found in Ref.~\cite{desjacques:ngsignature}, however these terms are second order in slow-roll parameters, therefore they are suppressed approximately by a factor $\mathcal{O}(\epsilon)$ with respect to the GE contribution. Furthermore we note that the first order correction to the Gaussian halo power spectra in equation~\eqref{eq:gaussian_halo_powerspectrum} and the Gaussian/non-Gaussian mixed contribution of equation~\eqref{eq:ps_mixed_contribution} become scale-independent at large scales, namely when taking the $k\to 0$ limit. This further highlights the fact that the GE scale dependence is quite unique, offering an opportunity to separate it from other signals. Moreover, as can be seen in equations~\eqref{eq:bispectrum_over_powerspectrum_scaling} and~\eqref{eq:trispectrum_over_powerspectrum_scaling}, the bispectrum contribution scales with redshift approximately as $(1+z)$ while for the trispectrum contribution the scaling is proportional to $(1+z)^2$; hence going to high redshift further helps the GE term to dominate over the bispectrum contributions, as can be explicitly seen in figure~\ref{fig:GE_in_PS}.

In conclusion, looking for this specific scale dependence at high redshift is a possible way to extract this specific signal from the halo power spectrum, providing an alternative way to determine the energy scale of inflation.


\subsection{Signal in the Halo Bispectrum}
\label{subsec:halo_bispectrum}
The Fourier transform of the Gaussian part of equation \eqref{eq:gaussian_34PF} reads as
\begin{equation}
\begin{aligned}
B^G_\mathrm{halo}(k_1, k_2, k_3, z) \approx& b^4_L(z) \left[P_R(k_1, z)P_R(k_2, z) + (2\ \mathrm{perms.}) \right]	\\
 +& b^6_L(z) \int \frac{d^3q}{(2\pi)^3} P_R(|\mathbf{k}_1-\mathbf{q}|, z)P_R(|\mathbf{k}_2-\mathbf{q}|, z)P_R(q, z)	\\
+ &\frac{b_L^6(z)}{2} \left[ P_R(k_1, z)\int\frac{d^3q}{(2\pi)^3}P_R(|\mathbf{k}_2-\mathbf{q}|, z)P_R(q, z) + (2\ \mathrm{perms.}) \right].
\end{aligned}
\label{eq:gaussian_halo_bispectrum}
\end{equation}
Even if the initial conditions are perfectly Gaussian,  we have a well-defined bispectrum of excursion regions. To compute the GE contribution to the halo bispectrum we take the Fourier transform of equation \eqref{eq:halo_threepoint_correlation}, obtaining
\begin{equation}
\begin{aligned}
B^{NG}_\mathrm{halo}(k_1, k_2, k_3, z) &\approx B^{G}_\mathrm{halo}(k_1, k_2, k_3, z) + B_\mathrm{123}(k_1, k_2, k_3, z) \\
&\quad + T_\mathrm{1123}(k_1, k_2, k_3, z) + T_\mathrm{1223}(k_1, k_2, k_3, z) + T_\mathrm{1233}(k_1, k_2, k_3, z)	\\
&\quad + M_{12-123}(k_1, k_2, k_3, z) + M_{23-123}(k_1, k_2, k_3, z) + M_{13-123}(k_1, k_2, k_3, z),
\end{aligned}
\label{eq:nongaussian_halo_bispectrum}
\end{equation}
where we recognise the non-Gaussian contributions,
\begin{equation}
\begin{aligned}
B_\mathrm{123}(k_1, k_2, k_3, z) &= b_L^3(z) \mathrm{FT}\left\lbrace\xi_R^{(3)}(\mathbf{x}_1, \mathbf{x}_2, \mathbf{x}_3)\right\rbrace\equiv   b_L^3(z) B_R(k_1,k_2,k_3)	\\
&= b_L^3(z) \mathcal{M}_R(k_1,z)\mathcal{M}_R(k_2,z)\mathcal{M}_R(k_3,z)B_\zeta(k_1, k_2, k_3),	\\
T_\mathrm{1123}(k_1, k_2, k_3, z) &= \frac{b_L^4(z)}{2} \mathrm{FT}\left\lbrace\xi_R^{(4)}(\mathbf{x}_1, \mathbf{x}_1, \mathbf{x}_2, \mathbf{x}_3)\right\rbrace	\\
&= \frac{b_L^4(z)}{2} \int \frac{d^3q}{(2\pi)^3}\mathcal{M}_R(q,z)\mathcal{M}_R(|\mathbf{k}_1-\mathbf{q}|,z)\mathcal{M}_R(k_2,z)\mathcal{M}_R(k_3,z) \times	\\
&\qquad\qquad\qquad\qquad\qquad\qquad \times T_\zeta(\mathbf{q}, \mathbf{k}_1-\mathbf{q}, \mathbf{k}_2, \mathbf{k}_{3}),	\\
T_\mathrm{1223}(k_1, k_2, k_3, z) &= \frac{b_L^4(z)}{2} \mathrm{FT}\left\lbrace\xi_R^{(4)}(\mathbf{x}_1, \mathbf{x}_2, \mathbf{x}_2, \mathbf{x}_3)\right\rbrace	\\
&= \frac{b_L^4(z)}{2} \int \frac{d^3q}{(2\pi)^3}\mathcal{M}_R(k_1,z)\mathcal{M}_R(|\mathbf{k}_2-\mathbf{q}|,z)\mathcal{M}_R(q,z)\mathcal{M}_R(k_3,z) \times	\\ &\qquad\qquad\qquad\qquad\qquad\qquad \times T_\zeta(\mathbf{k}_1, \mathbf{k}_2-\mathbf{q}, \mathbf{q}, \mathbf{k}_{3}),	\\
T_\mathrm{1233}(k_1, k_2, k_3, z) &= \frac{b_L^4(z)}{2} \mathrm{FT}\left\lbrace\xi_R^{(4)}(\mathbf{x}_1, \mathbf{x}_2, \mathbf{x}_3, \mathbf{x}_3)\right\rbrace	\\
&= \frac{b_L^4(z)}{2} \int \frac{d^3q}{(2\pi)^3}\mathcal{M}_R(k_1,z)\mathcal{M}_R(k_2,z)\mathcal{M}_R(q,z)\mathcal{M}_R(|\mathbf{k}_{3}-\mathbf{q}|,z) \times	\\
&\qquad\qquad\qquad\qquad\qquad\qquad \times T_\zeta(\mathbf{k}_1, \mathbf{k}_2, \mathbf{q}, \mathbf{k}_{3}-\mathbf{q}),
\end{aligned}
\end{equation}
and the mixed contributions,
\begin{equation}
\begin{aligned}
M_{12-123} + M_{23-123} + M_{13-123} = b_L^5(z) & \int\frac{d^3q}{(2\pi)^3} P_R(q,z) \Big[B_R(|\mathbf{k}_1-\mathbf{q}|,|\mathbf{k}_2+\mathbf{q}|,k_{12},z) + \Big. \\
&\qquad\qquad\qquad\quad + B_R(k_1,|\mathbf{k}_2-\mathbf{q}|,|\mathbf{k}_{12}-\mathbf{q}|,z) + \\
&\qquad\qquad\qquad\quad \Big. + B_R(|\mathbf{k}_1-\mathbf{q}|,k_2,|\mathbf{k}_{12}-\mathbf{q}|,z)\Big].
\end{aligned}
\label{eq:gaussian_nongaussian_mixedterm}
\end{equation}

\begin{figure}[h]
\centering
\includegraphics[width=\textwidth]{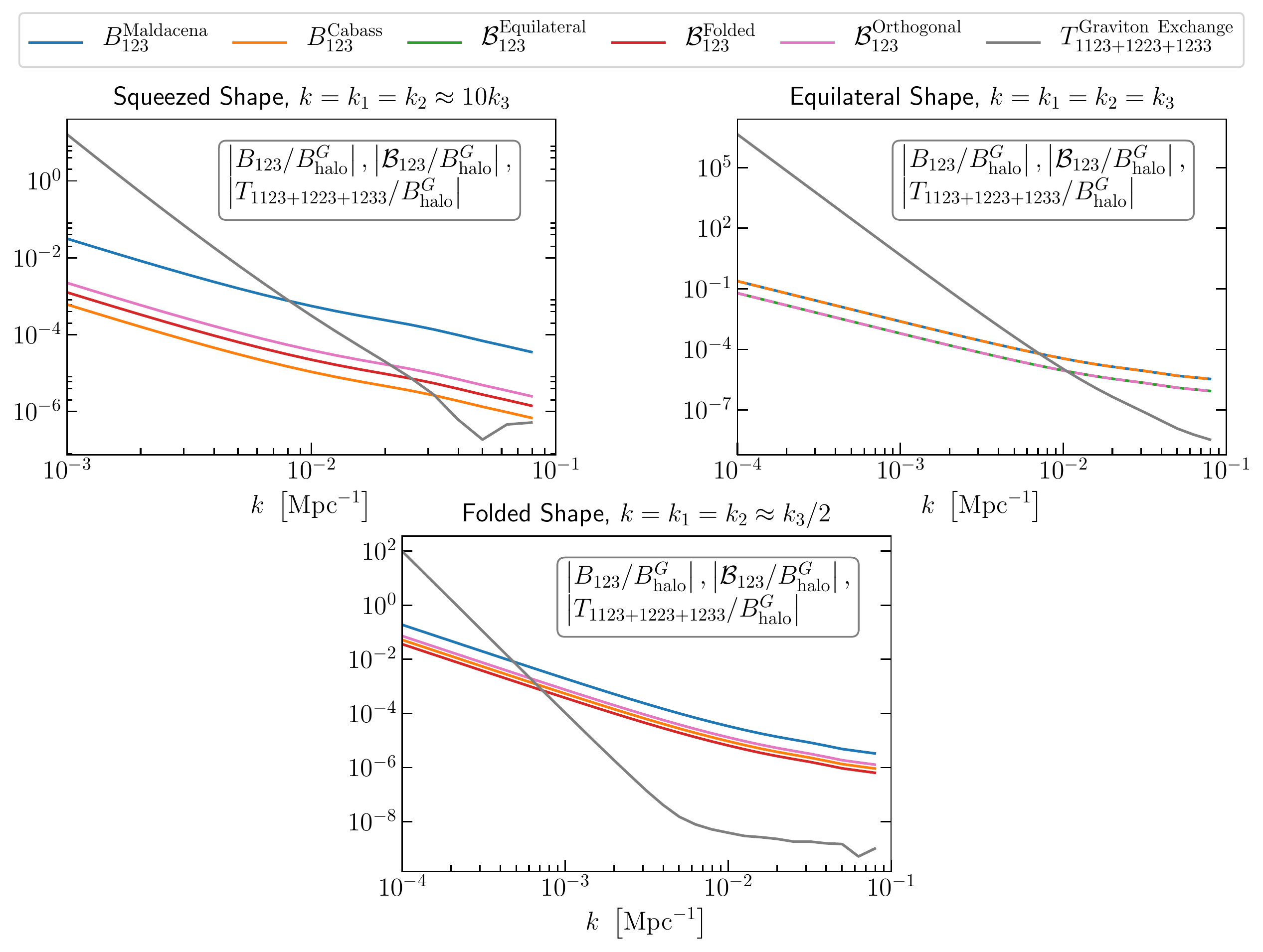}
\caption{Ratio between different primordial bispectra and GE trispectrum contribution with respect to the Gaussian halo bispectrum for squeezed (\textit{top left panel}), equilateral (\textit{top right panel}) and folded (\textit{bottom panel}) triangular shapes at redshift $z=0$ for $M_\mathrm{halo}=10^{12}\ M_\odot$ dark matter halos. We use $\epsilon=0.00625$ for Maldacena and Cabass bispectra, indicated by $B_{123}$, and $f_\mathrm{NL}=0.00279$ for the Equilateral, Folded and Orthogonal templates, indicated by $\mathcal{B}_{123}$. For the GE contribution we use $r=0.1$ and $k_\mathrm{hor}=10^{-6}\ \mathrm{Mpc}^{-1}$. Different values of $f_\mathrm{NL}$ and $r$ correspond to vertically scaling the Equilateral, Folded, Orthogonal templates and GE contribution, respectively.}
\label{fig:GE_in_BS}
\end{figure}

We compute the GE contribution following the same methodology described in the previous section, namely we substitute equation~\eqref{eq:graviton_exchange_trispectrum} into the four-point correlation function on the RHS of equation~\eqref{eq:nongaussian_halo_bispectrum}. As before, since we are interested only in primordial features, we report in figure~\ref{fig:GE_in_BS} the ratio between the primordial non-Gaussian contributions of equation~\eqref{eq:nongaussian_halo_bispectrum} and the Gaussian halo power spectrum. This allows us to compare the relative strength of the signals coming from Gaussian and non-Gaussian processes and the relative strength of the primordial bispectrum and trispectrum terms. 

In the three panels of figure \ref{fig:GE_in_BS}, since the exploration of every possible triangular configuration goes beyond the purpose of this work, we choose to explore just three representative triangular configuration, namely the equilateral ($k_1=k_2=k_3$), squeezed ($k_1=k_2\approx10k_3$) and folded ($k_1=k_2\approx k_3/2$) configurations. Also in this case we include, for comparison, different primordial bispectrum templates (see figure caption for the choice of normalisation). As seen also in section~\ref{subsec:halo_power_spectrum}, at large scales, in the case there is no primordial non-Gaussianity of any sort down to the ``gravitational floor'', the GE contribution easily dominates over the one arising from reasonably expected primordial non-Gaussian bispectrum. It is interesting to note that, at scales around $k\sim 10^{-3}\ \mathrm{Mpc}^{-1}$, the trispectrum contribution to the halo bispectrum in the squeezed and equilateral configurations becomes of the same order of the intrinsic halo bispectrum for an initial Gaussian field.
In the three panels we can identify the following scale and redshift scalings:
\begin{equation}
\begin{aligned}
&B_\mathrm{123}/B^G_\mathrm{halo},\ \mathcal{B}_\mathrm{123}/B^G_\mathrm{halo} \propto k^{-2}\frac{g(z)}{g(0)(1+z)},
\end{aligned}
\label{eq:bispectrum_over_bispectrum_scaling}
\end{equation}
which is valid for all models and templates except for those that vanish in specific triangular configurations, e.g., the Equilateral template in squeezed triangular configurations. On the other hand the trispectrum contributions scales as
\begin{equation}
\begin{aligned}
T_{1123+1223+1233}/B^G_\mathrm{halo} &\propto k^{-6},
\end{aligned}
\label{eq:trispectrum_over_bispectrum_scaling}
\end{equation}
independently from redshift, in contrast to the signal coming from primordial bispectra, which is suppressed approximately by a factor $(1+z)$ going to higher redshift. We do not report the magnitude of primordial bispectra signals in figure~\ref{fig:GE_in_BS} for redshift $z>0$, however the interested reader can simply divide the chosen model by the appropriate redshift factor, while keeping fixed the GE contribution, to get them. Since going to higher redshift shifts the primordial bispectra signal downward, the GE contribution will become even more dominant.

\begin{figure}[h]
\centering
\includegraphics[width=\textwidth]{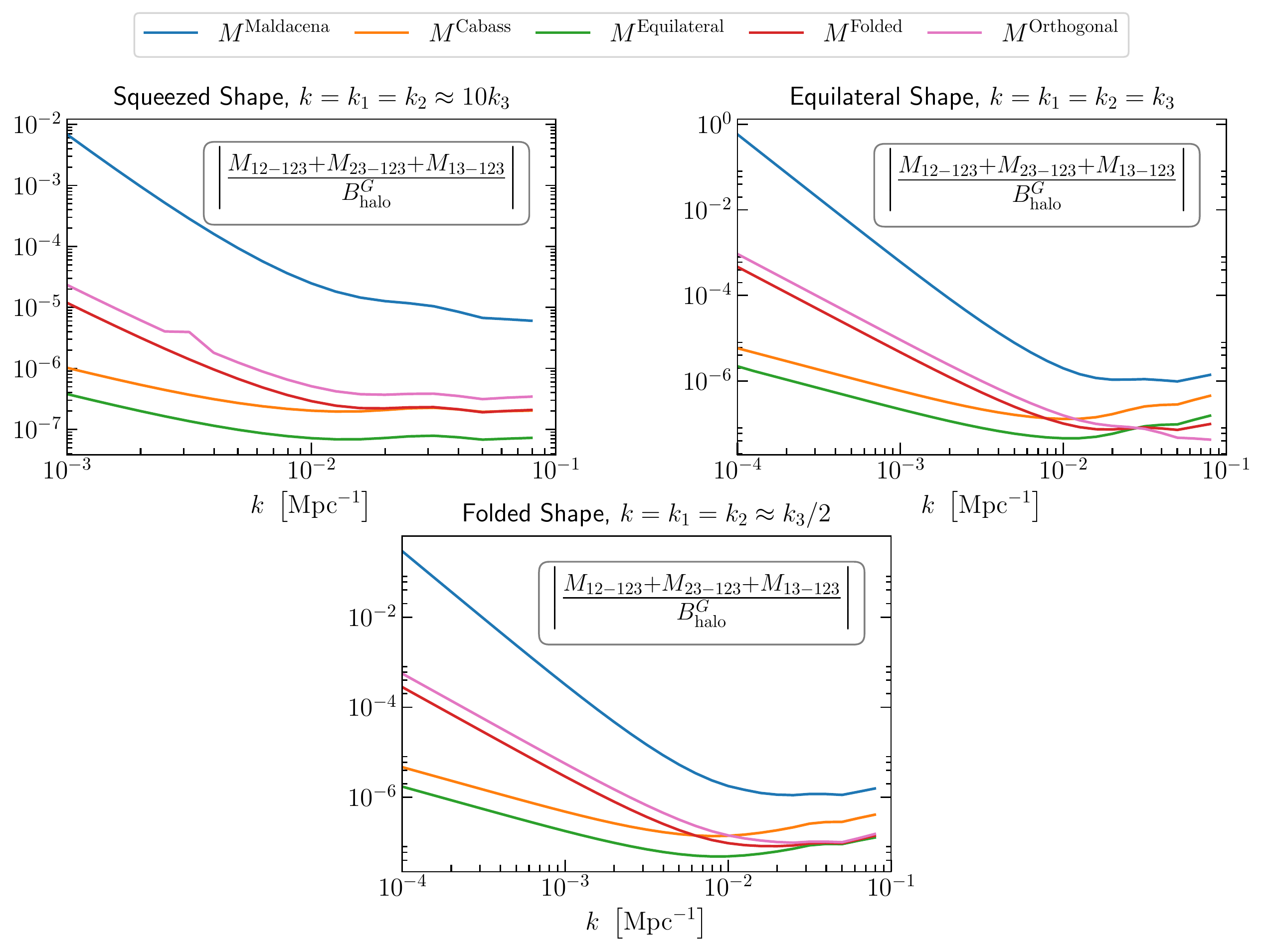}
\caption{Ratio between different Gaussian/non-Gaussian mixed terms with respect to the Gaussian halo bispectrum for squeezed (\textit{top left panel}), equilateral (\textit{top right panel}) and folded (\textit{bottom panel}) triangular shapes at redshift $z=0$ for $M_\mathrm{halo}=10^{12}\ M_\odot$ dark matter halos. We use $\epsilon=0.00625$ when Maldacena and Cabass bispectra appear in $M$, and $f_\mathrm{NL}=0.00279$ when the Equilateral, Folded and Orthogonal templates appear in the mixed term. Different values of $f_\mathrm{NL}$ correspond to vertically scaling the Equilateral, Folded, Orthogonal templates.}
\label{fig:BS_Mixed_Terms}
\end{figure}

Finally, we note that also in this case in all the configurations considered, Gaussian halo bispectrum corrections in equation~\eqref{eq:gaussian_halo_bispectrum} are scale-independent. On the other hand the mixed Gaussian/non-Gaussian term appearing in equation~\eqref{eq:gaussian_nongaussian_mixedterm} exhibits a potential scale dependence when taking the limit $k_1,k_2\to 0$. We report in figure~\ref{fig:BS_Mixed_Terms} the magnitude of this contribution relative to the Gaussian halo bispectrum at redshift $z=0$. As it can be seen from the figure, for our choice of parameters, the magnitude of this contribution is typically smaller than GE one, however, since this ratio grows approximately as $(1+z)$ with redshift, it might dominate over the GE signal at high redshift, depending on the real value of $r$ and~$f^\zeta_\mathrm{NL}$. Nevertheless its scale dependence is completely different from the characteristic one of the GE, therefore we still have some way to identify the signal we are interested in.

\section{Conclusions}
\label{sec:conclusions}
Determining the underlying physics of inflation is one of the big goals of Cosmology. A first step necessary to accomplish such a goal is determining the inflationary energy scale. In simple single-field slow-roll scenarios, the energy scale of inflation is proportional to the tensor-to-scalar ratio $r$ or, equivalently, to the first slow-roll parameter $\epsilon$. Several cosmological observables have been proposed  to measure the value of $r$, such as B-mode polarization and direct interferometric measurements of gravitational wave stochastic backgrounds.  In this work we explore a third avenue, the study 
of non-Gaussianities.

Non-Gaussianities are unavoidably produced during inflation and they constitute on their own a probe of the inflationary physics. Their importance as window into the self-interaction of the field during inflation is known (see e.g., Ref.~\cite{komatsu:ngprobeearlyuniverse} and references therein).  In this work we focused on the so-called graviton exchange, in particular on the specific non-Gaussianity generated by the interaction of scalar and tensor fluctuations at the horizon scale during the epoch of inflation. One of the peculiarities of this contribution to the four-point function is that it is suppressed only by one power of the slow-roll parameter. It becomes therefore interesting to entertain the idea that the GE contribution to the trispectrum  could be relevant for future large-scale galaxy surveys. Moreover, this avenue is worth exploring as the signal contains configurations that cannot be ``gauged'' away. This is not surprising as the graviton exchange is a real quantum effect and not an artefact due to local effects. 

We know from CMB observations that non-Gaussianities are small, in fact we have only upper bounds \cite{ade:planckng2013, ade:planckng2015, akrami:planckng2018}. Here we proposed to  look at   the $n$-point function of gravitationally collapsed structures to  further boost the signal coming from the primordial universe. In particular, we computed the contribution of the graviton exchange to the two- and three-point function of massive dark matter halos. We have shown that at large scales ($k \sim 10^{-4}-10^{-3}\ \mathrm{Mpc^{-1}}$) the contribution due to graviton exchange to the power spectrum of rare peaks is comparable to, if not dominant over,  the one  generated by the primordial three-point function expected from generic inflationary models (e.g., Maldacena  and Cabass bispectrum). We have also shown that this contribution has a particular scale dependence  and that it scales  with increasing  redshift faster than the three-point function contribution. Once going to high redshift favours the GE contributions  compared to other non-Gaussian signals. The same can be said to the GE contribution to the three-point function of dark matter halos for specifics configurations. This analytical approach to the clustering of peaks is of course an approximation to the clustering of realistic halos. While in detail the bias modelling for realistic halos may be much more complex than adopted here, the good agreement between simulations and the predictions obtained with this approach (see e.g., Refs.~\cite{dalal:simulations, desjacques:simulations, grossi:simulations, pillepich:simulations, nishimichi:simulations, wagner:bispectrumtemplatesI, wagner:bispectrumtemplatesII}) offers strong support that our initial investigation captures the behaviour of the signal both as a function of scale and redshift.

The effects produced by the GE contribution are significant at large scales, which are notoriously cosmic variance dominated. Since the signal depends on the tracer bias, the multi-tracer approach can be used beat down cosmic variance~\cite{seljak:cosmicvariance, mcdonald:cosmicvariance}. These results open an observational window, yet unexplored, but with the potential to help us  understand and verify the physics of inflation. This  new avenue  is highly complementary to direct or indirect (via CMB polarization) detection of primordial gravitational waves. We leave for future work a thorough computation of the observational configurations that have the largest signal-to-noise.


\acknowledgments
NBe. and LV acknowledge Martin Sloth and Filippo Vernizzi for helpful discussions. We thank Antonio Riotto for helpful comments.  NBe and LV  thank D. Baumann for an inspiring presentation at the ``Analytical Methods'' workshop of the Institut Henri Poincar\'e and thank the Center Emile Borel for hospitality during the latest stages of this work. Funding for this work was partially provided by the Spanish MINECO under projects AYA2014-58747-P AEI/FEDER, UE, and MDM-2014-0369 of ICCUB (Unidad de Excelencia Mar\'ia de Maeztu). NBe. is supported by the Spanish MINECO under grant BES-2015-073372. LV acknowledges support by European Union's Horizon 2020 research and innovation programme ERC (BePreSySe, grant agreement 725327). LV and RJ acknowledge the Radcliffe Institute for Advanced Study of Harvard University for hospitality during the latest stages of this work. NBa. and SM acknowledge partial financial support by ASI Grant No. 2016-24-H.0.

\bibliography{biblio}
\bibliographystyle{utcaps}


\appendix
\section{Bispectrum Templates}
\label{app:bispectrum_templates}
In general, the functional form of  the primordial bispectrum is complicated and  unsuitable for visualisation and data analysis. For this reason bispectrum {\it templates} have been constructed that are useful to  approximate the physical bispectrum and are suitable for data analysis. There is no shortage   of inflationary models where non-Gaussianities peak in configurations different from the squeezed one. In fact, if any of the conditions giving the standard, single-field, slow-roll is violated, important non-Gaussian signatures will be produced, and in particular the violation of each condition leaves its signature on specifics triangular configurations, see e.g., Ref.~\cite{komatsu:ngprobeearlyuniverse} and~\cite{chen:ngtriangularshapes} and Refs. therein. These types of non-Gaussianities, as shown in Ref.~\cite{senatore:nongaussianities}, are generically well described by a linear combination of three basic bispectrum templates. The widely known and used templates are the so-called, local, equilateral, folded and orthogonal. Of these four templates, only three are independent, the fourth  can obtained as a linear combination of the other tree see e.g., Refs.~\cite{senatore:nongaussianities, wagner:bispectrumtemplatesI, wagner:bispectrumtemplatesII}. For example the local template  is not independent from the other  three templates, in fact it can be described as a linear combination of them. Here below we report the most studied templates and  in the main text  we use them to check whether there is any particular shape that could contaminate the GE signal we are interested in. 

The equilateral template~\cite{babich:equilateralmodel}
\begin{equation}
\mathcal{B}_\zeta^\mathrm{Equilateral}(\mathbf{k}_1,\mathbf{k}_2,\mathbf{k}_3) = 6 f^\zeta_\mathrm{NL} \left(\frac{H^2_\star}{4\epsilon}\right)^2 \frac{\sum k_j^3}{\prod k_j^3}\left[- 1 + \frac{\sum_{i\neq j}k^2_ik_j - 2k_p}{\sum k_j^3}\right],
\label{eq:equilateral_bispectrum}
\end{equation}
is used to model non-Gaussianities arising from  e.g., inflaton Lagrangians with non-canonical kinetic terms; in this case  the bispectrum is peaked on equilateral shapes. 

The folded template~\cite{chen:foldedmodel1, chen:foldedmodel2, holman:foldedmodel, meerburg:foldedmodel}
\begin{equation}
\mathcal{B}_\zeta^\mathrm{Folded}(\mathbf{k}_1,\mathbf{k}_2,\mathbf{k}_3) = 6 f^\zeta_\mathrm{NL} \left(\frac{H^2_\star}{4\epsilon}\right)^2 \frac{\sum k_j^3}{\prod k_j^3}\left[ 1 + \frac{3k_p - \sum_{i\neq j}k^2_ik_j}{\sum k_j^3}\right],
\label{eq:folded_bispectrum}
\end{equation}
 is used to model non-gaussianities arising  from different assumption on the initial vacuum state. 
 
The orthogonal template~\cite{senatore:nongaussianities}
\begin{equation}
\mathcal{B}_\zeta^\mathrm{Orthogonal}(\mathbf{k}_1,\mathbf{k}_2,\mathbf{k}_3) = 6 f^\zeta_\mathrm{NL} \left(\frac{H^2_\star}{4\epsilon}\right)^2 \frac{\sum k_j^3}{\prod k_j^3}\left[- 3 + \frac{3\sum_{i\neq j}k^2_ik_j - 8k_p}{\sum k_j^3}\right],
\label{eq:orthogonal_bispectrum}
\end{equation}
where $k_p=\prod_{j=1}^{3} k_j$ is the product of the three momenta, has been built to be orthogonal to the equilateral one.

\end{document}